# Picosecond-petawatt laser-block ignition for avalanche fusion of boron by ultrahigh acceleration and ultrahigh magnetic fields

**H Hora**[*1], **P Lalousis**[2], **L Giuffrida**[3], **D Margarone**[3], **G Korn**[3], **S Eliezer**[4,5], **G H Miley**[6], **S Moustaizis**[7], **G Mourou**[8] **and C P J Barty**[9]

[1]Department of Theoretical Physics, University of New South Wales, Sydney 2052, Australia
[2]Institute of Electronic Structure and Lasers FORTH, Heraklion, Greece
[3]ELI-Beamline Project, Inst. Physics, ASCR, PALS Center, Prague, Czech Republic
[4]Institute of Nuclear Fusion, Polytechnic University of Madrid, Madrid, Spain
[5]Soreq Research Center, Yavne, Israel
[6]Deptartment of Nucluclear Plasma & Radiological Engineering University of Illinois, Urbana IL, USA,
[7] Technical University Crete, Laboratory of Matter Structure and Laser Physics, Chania, Greece
[8] DAER-IZEST, Ecole Polytechnique, Route de Saclay, 91128 Palaiseau Cedex, France
[9]Lawrence Livermore National Laboratory, Livermore, CA, USA
*h.hora@unsw.edu.au

**Abstract.** Fusion energy from reacting hydrogen (protons) with the boron isotope 11 (HB11) resulting in three stable helium nuclei, is without problem of nuclear radiation in contrast to DT fusion. But the HB11 reaction driven by nanosecond laser pulses with thermal compression and ignition by lasers is extremely difficult. This changed radically when irradiation with picosecond laser pulses produces a non-thermal plasma block ignition with ultrahigh acceleration. This uses the nonlinear (ponderomotive) force to surprizingly resulting in same thresholds as DT fusion even under pessimistic assumption of binary reactions. After evaluation of reactions trapped cylindrically by kilotesla magnetic fields and using the measured highly increased HB11 fusion gains for the proof of an avalanche of the three alphas in secondary reactions, possibilities for an absolutely clean energy source at comptitive costs were concluded.

## 1. Introduction

Energy generation from laser driven fusion of deuterium D with tritium T (DT) arrived at highest fusion gains at the NIF project [1][2] by using indirect drive using up to 2 MJ laser pulses of about nanosecond duration where the thermal processes with ablation-compression and spark ignition are involved. A similar option with direct drive and volume ignition is being studied [3] for thermal driven laser-fusion with nanosecond laser pulses.

   This is in basic contrast with using picosecond (ps) laser pulses with needing energy fluxes E* of about $10^8$ J/cm$^2$ for igniting uncompressed solid density DT. Initiated by Chu [4], a non-thermal energy transfer from laser energy into plasma blocks is used to avoid complicate thermal determined mechanisms of the nanosecond interaction. This is possible now through the Chirped Pulse Amplification CPA [5] of laser pulses of ps and shorter duration with powers above petawatt (PW). The ultrahigh acceleration of the plasma blocks was theoretically-numerically predicted in 1978 and measured [6] in agreement with the theory [7].

   The advantage of the ps-block-ignition arrived at the surprising result [8][9], that the environmentally clean fusion of protons with the boron isotope 11 (HB11) is possible at similar thresholds as DT, while thermal ignition was considered as impossible with needing compression of HB11 to more than 100000 times solid state density. Next steps of these developments focus on the combination of these results with cylindrical trapping of the fusion reaction with ultrahigh magnetic fields of about 10 kilotesla and by using the measured highly increased gains of HB11 fusion as an avalanche ignition process.

## 2. Block ignition by picosecond laser pulses

The drastic difference between the interactions of laser pulses of ps and ns duration is given by the force density **f** in the plasma being not only determined by the gas dynamic pressure p but also by the force $\mathbf{f}_{NL}$ due to electric **E** and magnetic **B** laser fields of frequency ω,

$$\mathbf{f} = -\nabla p + \mathbf{f}_{NL} \tag{1}$$

where the force $\mathbf{f}_{NL}$ is given by Maxwell's stress tensor as Lorentz and gauge invariant nonlinear force

$$\mathbf{f}_{NL} = \nabla \bullet [\mathbf{EE} + \mathbf{HH} - 0.5(\mathbf{E}^2 + \mathbf{H}^2)\mathbf{1} + (1+(\partial/\partial t)/\omega)(\mathbf{n}^2-1)\mathbf{EE}]/(4\pi) - (\partial/\partial t)\mathbf{E}\times\mathbf{H}/(4\pi c) \tag{2}$$

where **1** is the unity tensor and **n** is the complex optical constant of the plasma given by the plasma frequency $\omega_p$. At plane laser wave interaction with a plane plasma front, the nonlinear force reduces to

$$\mathbf{f}_{NL} = -(\partial/\partial x)(\mathbf{E}^2+\mathbf{H}^2)/(8\pi) = -(\omega_p/\omega)^2(\partial/\partial x)(E_v^2/\mathbf{n})/(16\pi) \tag{3}$$

showing how the force density is given by the negative gradient of the electromagnetic laser-field energy-density including the magnetic laser field from Maxwell's equations. $E_v$ is the amplitude of the electric laser field in vacuum after time averaging. The second expression in Eq. (3) is Kelvin's formulation of the ponderomotive force in electrostatics of 1845 [10].

The difference to laser interaction by ns thermal interaction against ps non-thermal nonlinear force driving is determined by $\mathbf{f}_{NL}$ interaction dominating in Eq. (1). For the ns interaction, the first term dominates at low laser intensities while with ps, the second term dominates in which case the quiver energy of the electrons of the laser field has to be higher than their thermal energy of motion. A numerical example about nonlinear force acceleration of a slab of deuterium plasma irradiated by a neodymium glass laser pulse is shown in Figure 1. During the 1.5 ps, the plasma reached velocities above $10^9$ cm/s by the ultrahigh acceleration above $10^{20}$ cm/s$^2$. The generation of the plasma blocks, one moving against the laser light and the other into the higher density target is the result of a non-thermal collisionless absoprtion and should not be understood as radiation pressure acceleration but as a dielectric explosion driving the plasma blocks.

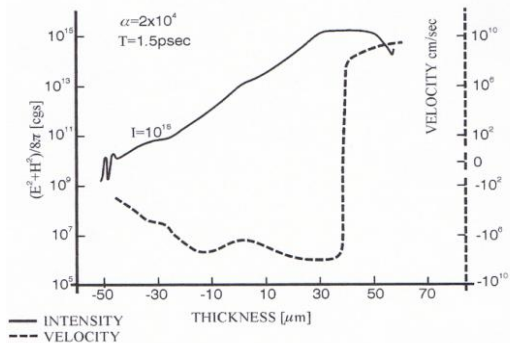

Fig. 1. $10^{18}$ W/cm$^2$ neodymium glass laser incident from the right hand side on an initially 100 eV hot deuterium plasma slab whose initial density has a very low reflecting bi-Rayleigh profile, resulting in a laser energy density and a velocity distribution from plasma hydrodynamic computations at time t=1.5 ps of interaction. The driving nonlinear force is the negative of the energy density gradient of the laser field $(\mathbf{E}^2+\mathbf{H}^2)/8\pi$. The dynamic development of temperature and density had accelerated the plasma block of about 15 vacuum wave length thickness of the dielectric enlarged skin layer moving against the laser (positive velocity) and another block into the plasma (negative velocity) showing ultrahigh $>10^{20}$ cm/s$^2$ acceleration ([10]: figures 4.14 & 4.15) as computer result of 1978.

The experimental proof of the ultrahigh acceleration was possible [6] in full agreement with the results of computations in 1978, after laser pulses of higher than terawatt (TW) power and about ps duration were available only thanks to the Chirped Pulse Amplification CPA [5]. With these ps ultrahigh accelerations the plasma block ignition of solid density DT by the nonlinear force was possible and updated [8]. Computation of DT and HB11 fusion using the ps-block ignition [8] showed many details of the generated fusion flames with velocities of few 1000 km/s, the delayed generation of a Rankine-Hugoniot shock fronts, local distribution of reaction rates etc. [11], however, in one dimension plane wave computations. Using spherical laser irradiation needed laser powers in the exawatt range for gains of few hundred.

### 3. Secondary avalanche reactions

The results reported up to this stage were based on calculations of binary reactions as it is the case for DT. These binay reactions for protons with $^{11}$B used in the computations, however, were a rather pessimistic assumption. The three 2.9 MeV alphas from an initial binary reaction can by elastic collisions transfer energy to boron nuclei or protons such that they secondarily produce each three further alphas with avalanching to a high energy gain. Computations were performed [13] confirming the secondary reactions.

Before considering the elastic collisions in view of stoping power and related processes it was fortunate that experiments became known [12] from which the avalanche process was evident [13]. Following the groundbreaking measurements of HB11 reactions of more than $10^6$ under rather complicate experimental conditions [14], the transparent conditions of the measurement of $10^9$ reactions per steradian [12] by the PALS iodine laser based on few hundred Joules laser pulses in the range of 100 ps duration were so strongly elevated even above comparable results with DT [13] that this could only be due to the secondary reactions with the avalanche.

The avalanche HB11 reaction can be explained in quantitative agreement with the experiment [12] numerically by evaluating an extremely non-equilibrium state of the plasma by elastic collisions of the alpha particles with the boron nuclei and protons where the broad energy range around 600keV fits with the most exceptionally increased HB11 fusion cross section [15].

Based on this clarification for the avalanche secondary processes, the following cylindrical trapping of the HB11 reaction with solid density fuel could be used [16] where the ultrahigh magnetic fields in the range of 10 kilotesla [17] were applied, Figure 2.

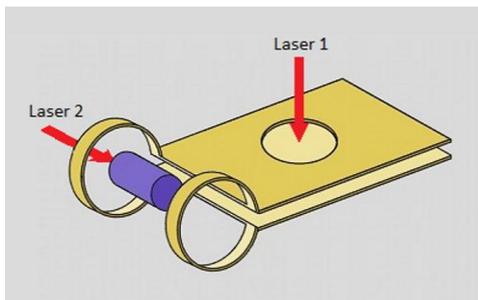

*Figure 2. Generation of a 10 kiloTesla magnetic field of about 2ns duration in the coils by firing a >kilojoule nanosecond laser pulse 1 into the hole between the plates. The HB11 fusion fuel is a solid cylinder of 1cm length and 1 mm radius coaxially located in the coils and the block-ignition of the fusion flame is produced by a ps-30 PW laser pulse of 0.2 mm diameter for block ignition from laser 2.*

Based on the computation [16] of block ignition by a ps laser pulse 2 in Figure 2 of $10^{20}$W/cm$^2$ intensity on a solid HB11 cylinder of 0.2mm diameter within a 10 kilotesla magnetic field shows a slow radial expansion against the trapping and an axial propagation of the reaction with several thousands of km/s velocity during about a ns. With avalanche reaction, the cylinder of Fig. 2 is nearly completely reacting producing more than one gigajoule energy of alpha particles by the irradiated 30 kJ laser pulse 2.

### 4. Estimations about preliminary parameters for HB11 fusion generation.

Following the results [16] for a reaction scheme of Fig. 2, a ps laser pulse of 30 kilojoule energy (30 PW power) should produce more than a GJ energy in clean alpha particles. The technology for the needed laser pulses is close to availability in view of 200 PW-0.1ps laser pulses to be produced using the developed NIF-technology [18]. Figure 3 describes the scheme of a fusion reactor for converting the energy of the alpha particles into an electrostatic pulse in the range close to 1.4 megavolts using the technology of high voltage direct current HVDC power transmission electricity.

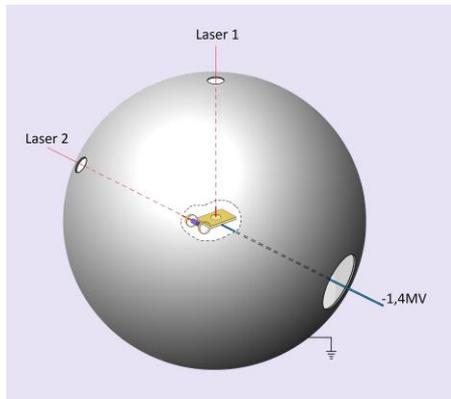

*Figure 3. Scheme of a HB11 fusion reactor without any radioactive radiativon problems is based on non-thermal plasma block ignition by nonlinear forces using a 30kJ-picosecond laser pulse 2. The central reaction unit (Figure 2) located in the center of the reactor sphere is electric charged to the level of -1.4 million volts against the wall of the sphere such that the alpha particles (helium nuclei) produce more than a gigajoule energy, of which a small part is needed for the operation of the laser pulses. One part of the gained costs of electricity is needed for the reaction unit being destroyed and the HB11 fuel at each reaction [13][15].*

It was estimated [13] for a power station working with one reaction per second, that after deducting the costs for investment and operation, the energy of the value of up to $300Million per year may be produced for the grid.